\documentclass[amsmath,aps,showpacs,a4paper,10pt]{revtex4}

 \usepackage{epsf}
 \usepackage{graphicx}    

 \usepackage{enumerate}   

\begin{document}

 \newcommand{\be}[1]{\begin{equation}\label{#1}}
 \newcommand{\ee}{\end{equation}}
 \newcommand{\bea}{\begin{eqnarray}}
 \newcommand{\eea}{\end{eqnarray}}
 \def\disp{\displaystyle}

 \def\gsim{ \lower .75ex \hbox{$\sim$} \llap{\raise .27ex \hbox{$>$}} }
 \def\lsim{ \lower .75ex \hbox{$\sim$} \llap{\raise .27ex \hbox{$<$}} }

 \begin{titlepage}

 \begin{flushright}
 arXiv:1808.00377
 \end{flushright}

 \title{\Large \bf Non-parametric Reconstruction of Growth
 Index via~Gaussian~Processes}

 \author{Zhao-Yu~Yin\,}
 \email[\,email address:\ ]{854587902@qq.com}
 \affiliation{School of Physics,
 Beijing Institute of Technology, Beijing 100081, China}

 \author{Hao~Wei\,}
 \email[\,Corresponding author;\ email address:\ ]{haowei@bit.edu.cn}
 \affiliation{School of Physics,
 Beijing Institute of Technology, Beijing 100081, China}

 \begin{abstract}\vspace{1cm}
 \centerline{\bf ABSTRACT}\vspace{2mm}
 The accelerated cosmic expansion could be due to dark energy
 within general relativity (GR), or modified gravity. It is of
 interest to differentiate between them, by using both the expansion
 history and the growth history. In the literature, it was proposed
 that the growth index $\gamma$ is useful to distinguish these
 two scenarios. In this work, we consider the non-parametric
 reconstruction of the growth index $\gamma$ as a function of
 redshift $z$ from the latest observational data as of July 2018 via
 Gaussian Processes. We find that $f(R)$ theories and dark
 energy models within GR (especially $\Lambda$CDM) are inconsistent
 with the results in the moderate redshift range far beyond
 $3\sigma$ confidence level. A modified gravity scenario
 different from $f(R)$ theories is favored. However, these
 results can also be due to other non-trivial possibilities,
 in which dark energy models within GR (especially $\Lambda$CDM) and
 $f(R)$ theories might still survive. In all cases, our
 results suggest that new physics is required.
 \end{abstract}


 \maketitle

 \end{titlepage}

 \renewcommand{\baselinestretch}{1.0}


\section{Introduction}\label{sec1}

Since the discovery of the accelerated expansion of our
 universe in 1998~\cite{Riess:1998cb,Perlmutter:1998np}, the
 real cause of this mysterious phenomenon is still unclear
 so far. As is well known, two main types of scenarios are
 extensively considered in the literature to this end. The first one
 is to introduce an unknown component with negative pressure
 (dark energy) in the framework of general relativity (GR). On
 the contrary, the second one explains the accelerated
 expansion by using a modification to GR (modified gravity),
 without invoking dark energy. We refer to
 e.g.~\cite{Ishak:2018his,Amendola:2016saw,Joyce:2014kja,
 Clifton:2011jh} for comprehensive reviews.

Until now, both scenarios are competent to interpret the
 accelerated cosmic expansion. Therefore, it is of interest to
 differentiate between them. Since they cannot be distinguished
 by using the expansion history solely, it is necessary to consider
 the growth history in addition (see e.g.~\cite{Linder:2005in,
 Linder:2007hg,Wei:2008ig,Wei:2008vw} and references therein).
 In fact, if the models of dark energy and modified gravity share a
 same expansion history, they might have different growth histories.
 Typically, the growth history is characterized by the linear matter
 density contrast $\delta(z)\equiv\delta\rho_m/\rho_m$ as a function
 of redshift $z$. It is convenient to introduce the growth rate
 $f\equiv d\ln\delta/d\ln a$, where $a=(1+z)^{-1}$ is the scale
 factor. Many years ago, a good approximation $f=\Omega_m^\gamma$ has
 been first proposed in~\cite{Peebles1980,Lahav:1991wc} within
 GR, where $\gamma$ is the growth index, and $\Omega_m$ is
 the fractional density of pressureless matter.
 In the beginning, $f=\Omega_m^\gamma$ was used only at the
 present time ($z=0$), and it was not valid for any redshift.
 Since~\cite{Wang:1998gt} it was applied to anything beyond
 matter, curvature, and a cosmological constant. Finally,
 not until~\cite{Lue:2004rj} was it applied to gravity other
 than GR, and then in~\cite{Linder:2005in} generalized to
 modified gravity, varying equation of state, and an integral
 relation for growth. Nowadays, the general form
 $f(z)=\Omega_m(z)^{\gamma(z)}$ has been extensively used
 in the literature.

In e.g.~\cite{Linder:2005in,Linder:2007hg}, it was proposed that the
 growth index $\gamma$ is useful to distinguish the scenarios
 of dark energy and modified gravity. In GR,
 $\gamma=6/11\simeq 0.545$ for $\Lambda$CDM
 model~\cite{Linder:2005in,Linder:2007hg} (which
 is approximately independent of redshift), while $\gamma\simeq 0.55$
 for many dark energy models~\cite{Linder:2005in}. In the cases
 of modified gravity, $\gamma\simeq 0.68$ for Dvali-Gabadadze-Porrati
 (DGP) model ($\gamma=11/16$ is its high redshift asymptotic
 value)~\cite{Linder:2007hg,Wei:2008ig}, while
 $\gamma\simeq 0.42$ for most of viable $f(R)$ theories
 ($\gamma\,\lsim\, 0.557$ certainly for almost all viable
 $f(R)$ theories, and $\gamma$ decreases when redshift
 increases)~\cite{Gannouji:2008wt,Tsujikawa:2009ku,Shafieloo:2012ms,
 Tsujikawa:2010zza}. Since their $\gamma(z)$ lie in a narrow
 range around the above values respectively, one
 might differentiate between them.

In the literature, the growth indices for some particular models have
 been constrained by using the observational data, but only
 the present value $\gamma_0$ and the derivative
 $\gamma^\prime_0$ were considered usually. Of course, it is better
 to study the growth index in a model-independent way. In the
 literature, a common choice is to consider the
 model-independent parameterizations for $\gamma(z)$, but a
 particular function form should be given {\it a~prior}. On the
 contrary, it is worth noting that the goal function could be
 directly reconstructed from the input data by using some
 non-parametric methods, such as principal component analysis,
 and Gaussian processes, without assuming a particular function form.

Here, we consider the non-parametric reconstruction of the
 growth index $\gamma(z)$ as a function of redshift~$z$ via Gaussian
 processes~\cite{Rasmussen:2006,Seikel:2012uu}, by using the
 latest observational data. In Sec.~\ref{sec2}, we briefly
 describe the methodology. In Secs.~\ref{sec3} and \ref{sec4},
 the results and the conclusions are given, respectively. We
 find that $f(R)$ theories, and dark energy models within GR
 (especially $\Lambda$CDM), are inconsistent with the results in the
 moderate redshift range, far beyond $3\sigma$ confidence level
 (C.L.). A modified gravity scenario different from $f(R)$ theories
 is favored. However, there might be other possibilities for
 these results, and we will discuss this issue briefly in
 Sec.~\ref{sec4}. In all cases, our results suggest that new
 physics is required.


 \begin{center}
 \begin{figure}[tb]
 \centering
 \vspace{-6mm}  
 \includegraphics[width=0.85\textwidth]{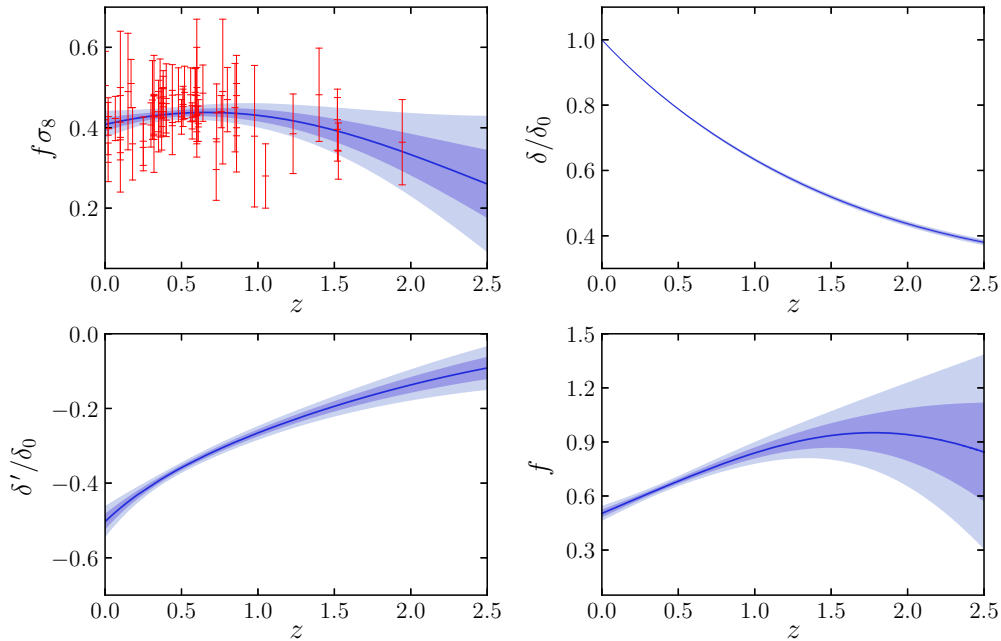}
 \caption{\label{fig1} The reconstructed $f\sigma_8$, $\delta/\delta_0$,
 $\delta^\prime/\delta_0$ and $f$ as functions of redshift $z$,
 by using Gaussian processes with the squared exponential
 covariance function. The mean and $1\sigma$, $2\sigma$
 uncertainties are indicated by the blue solid lines and the
 shaded regions, respectively. The observational $f\sigma_{8,\,obs}$
 data with error bars are also plotted in the top-left
 panel. See the text for details.}
 \end{figure}
 \end{center}


\vspace{-12mm}  


\section{Methodology}\label{sec2}

By definition $f(z)=\Omega_m(z)^{\gamma(z)}$, the growth
 index is given by
 \be{eq1}
 \gamma(z)=\frac{\ln f(z)}{\ln\Omega_m(z)}\,.
 \ee
 Note that in a few works (e.g.~\cite{Gonzalez:2016lur,
 Gonzalez:2017tcm}) a fairly
 different $\gamma(z)=d\ln f(z)/d\ln\Omega_m(z)$ is taken,
 which coincides with Eq.~(\ref{eq1}) only when
 $\gamma=const.$ To reconstruct $\gamma(z)$, both $f(z)$
 and $\Omega_m(z)$ are needed.

The growth rate $f$ can be obtained from redshift space
 distortion (RSD) measurements, and the observational $f_{obs}$
 data have been used in some relevant works (e.g.~\cite{Wei:2008ig,
 Gonzalez:2016lur,Gonzalez:2017tcm}). However, it is sensitive
 to the bias parameter $b$ which can vary in the range
 $b\in[1,\,3]$. This makes the observational $f_{obs}$ data
 unreliable~\cite{Nesseris:2017vor}. Instead, the combination
 $f\sigma_8(z)\equiv f(z)\,\sigma_8(z)$ is independent of the
 bias, and hence is more reliable, where $\sigma_8(z)=\sigma_8(z=0)
 \,\delta(z)/\delta(z=0)=\sigma_{8,\,0}\,\delta(z)/\delta_0$ within
 spheres of radius $8h^{-1}$Mpc~\cite{Nesseris:2017vor}, and
 the subscript ``0'' indicates the present value
 of the corresponding quantity. Noting that
 \be{eq2}
 f\equiv\frac{d\ln\delta}{d\ln a}
 =-(1+z)\,\frac{\delta^\prime}{\delta}\,,
 \ee
 where a prime denotes the derivative with
 respect to redshift $z$, we have
 \bea
 \frac{\delta^\prime(z)}{\delta_0}&=&
 -\frac{1}{\sigma_{8,\,0}}\frac{f\sigma_8(z)}{1+z}\,,\label{eq3}\\[1mm]
 \frac{\delta(z)}{\delta_0} &=& 1-\frac{1}{\sigma_{8,\,0}} \int_0^z
 \frac{f\sigma_8(\tilde{z})}{1+\tilde{z}}\,d\tilde{z}\,.\label{eq4}
 \eea
 In fact, the observational $f\sigma_{8,\,obs}$ data can be obtained
 from weak lensing and RSD measurements~\cite{Nesseris:2017vor,
 Kazantzidis:2018rnb}. Once $f\sigma_8(z)$ is reconstructed
 from the observational $f\sigma_{8,\,obs}$ data via Gaussian
 processes~\cite{Rasmussen:2006,Seikel:2012uu}, we can obtain
 $\delta^\prime(z)/\delta_0$, $\delta(z)/\delta_0$, and finally
 $f(z)$ by using Eqs.~(\ref{eq3}), (\ref{eq4}) and (\ref{eq2}).


 \begin{center}
 \begin{figure}[tb]
 \centering
 \vspace{-6mm}  
 \includegraphics[width=0.85\textwidth]{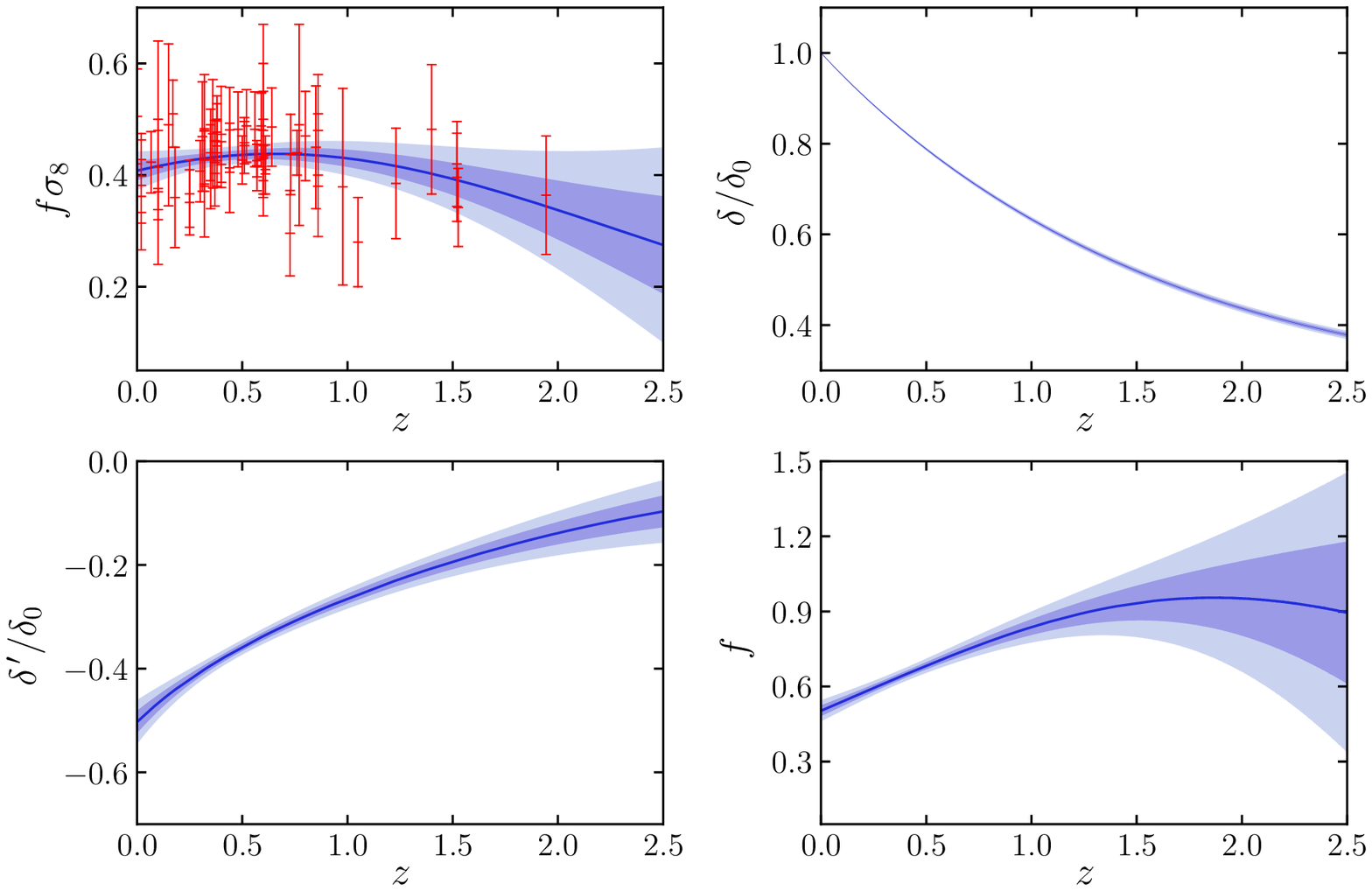}
 \caption{\label{fig2} The same as in Fig.~\ref{fig1}, except
 for the Mat\'ern ($\nu=9/2$) covariance function. See the
 text for details.}
 \end{figure}
 \end{center}


\vspace{-10.7mm} 

On the other hand, the dimensionless Hubble parameter $E\equiv H/H_0$
 is required to reconstruct
 \be{eq5}
 \Omega_m(z)\equiv\frac{8\pi G\rho_m}{3H^2}=
 \frac{\Omega_{m0}(1+z)^3}{E^2(z)}\,.
 \ee
 Note that $E(z)$ can also be obtained from the growth
 history~\cite{Chiba:2007rb,Starobinsky:1998fr,Zhang:2018gjb}.
 In GR, the perturbation equation
 $\ddot{\delta}+2H\dot{\delta}=4\pi G\rho_m\delta$ can be recast as
 a differential equation for $H^2$, where a dot denotes the
 derivative with respect to cosmic time $t$. Its solution is
 given by~\cite{Chiba:2007rb,Starobinsky:1998fr,Zhang:2018gjb}
 \be{eq6}
 E^2=3\Omega_{m0}\,\frac{(1+z)^2}
 {\delta^{\prime\,2}}\int_z^\infty\frac{\delta}{1+\tilde{z}}
 \left(-\delta^\prime\right)d\tilde{z}\,.
 \ee
 However, in the case of modified gravity, the perturbation equation
 becomes $\ddot{\delta}+2H\dot{\delta}=4\pi G_{\rm eff}\rho_m\delta$,
 and its solution reads~\cite{Chiba:2007rb}
 \be{eq7}
 E^2=3\Omega_{m0}\,\frac{(1+z)^2}
 {\delta^{\prime\,2}}\int_z^\infty\frac{-\delta\delta^\prime}
 {1+\tilde{z}}\cdot\frac{G_{\rm eff}}{G}\,d\tilde{z}\,,
 \ee
 where $G_{\rm eff}=G\,(1+1/(3\beta))$ is the effective gravitational
 ``constant'', and $\beta$ depends on time in general, which will be
 known if the modified gravity is specified. For example,
 $\beta=-(1+\Omega_m^2)/(1-\Omega_m^2)$ for the flat DGP
 model~\cite{Chiba:2007rb,Wei:2008ig,Wei:2008vw}. Noting
 Eq.~(\ref{eq5}), it is difficult to analytically obtain $E(z)$ from
 Eq.~(\ref{eq7}) because $E^2$ appears in the both sides. On
 the other hand, if we do not know whether GR is modified or
 not {\it a~prior} (since our goal is to differentiate between
 dark energy within GR, and modified gravity), we also do not
 know which one of Eqs.~(\ref{eq6}) and (\ref{eq7}) will be
 used. Therefore, it is not viable to obtain $E(z)$ by using
 the growth history.

The only viable way is to use the expansion history. There exist two
 different approaches to this~end. The first one is to directly
 reconstruct $E(z)$ by using the observational
 $H(z)$ data~\cite{Magana:2017nfs,Geng:2018pxk,Wei:2006ut} from
 the measurements of the differential age and the baryon
 acoustic oscillation (BAO). The second one is to
 use the luminosity distance of type Ia supernovae (SNIa),
 $d_L(z)=(c/H_0)(1+z)\,D(z)$, where $c$ is the speed of light.
 Note that we consider a flat Friedmann-Robertson-Walker (FRW)
 universe throughout. In this case,
 $D=\int_0^z d\tilde{z}/E(\tilde{z})$, and hence $E=1/D^\prime$. Once
 $E(z)$ is reconstructed from the observational $H(z)$ data or
 SNIa via Gaussian processes, we can obtain $\Omega_m(z)$ by
 using Eq.~(\ref{eq5}). Finally, the growth index $\gamma(z)$
 is available from Eq.~(\ref{eq1}).


 \begin{center}
 \begin{figure}[tb]
 \centering
 \vspace{-7mm}  
 \includegraphics[width=0.85\textwidth]{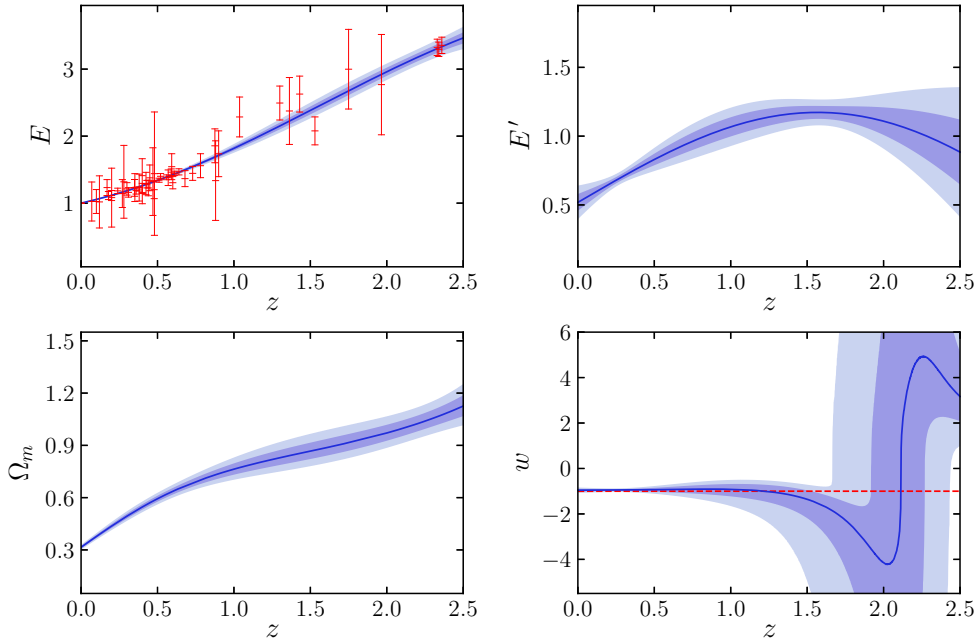}
 \caption{\label{fig3} The reconstructed $E$, $E^\prime$,
 $\Omega_m$ and $w$ as functions of redshift $z$,
 by using Gaussian processes with the squared exponential
 covariance function, from the observational $H(z)$ data with
 $H_0=67.36\pm 0.54\,$km/s/Mpc. The mean and $1\sigma$,
 $2\sigma$ uncertainties are indicated by the blue solid lines
 and the shaded regions, respectively. The observational
 $E_{obs}$ data with error bars are also plotted
 in the top-left panel. $w=-1$ is indicated by a red dashed
 line. See the text for details.}
 \end{figure}
 \end{center}


\vspace{-10.2mm} 

We refer to e.g.~\cite{Rasmussen:2006,Seikel:2012uu,Seikel:2013fda}
 for the details of Gaussian processes. In this work, we
 implement Gaussian processes by using the code GaPP (Gaussian
 Processes in Python)~\cite{Seikel:2012uu}. In Gaussian
 processes, there are many options for the covariance function
 (or, the kernel function) $\kappa (z,\,\bar{z})$. Here, we
 choose to use two different types of $\kappa (z,\,\bar{z})$.
 The first one is the squared exponential (or, Gaussian) covariance
 function, which is the simplest and popular choice in the
 literature. The second one is the Mat\'ern ($\nu=9/2$)
 covariance function, which is recommended by~\cite{Seikel:2013fda}
 because it is the best in the ones under consideration. The
 explicit forms of these two covariance functions can be found
 in e.g.~\cite{Rasmussen:2006,Seikel:2012uu,Seikel:2013fda,
 Zhang:2018gjb,Cai:2016sby}.

Finally, it is of interest to fully extract information from
 the expansion history. In modified gravity, the modification
 to GR can also be regarded as an ``effective dark energy''
 component in GR. The equation-of-state parameter (EoS) of
 the real/effective dark energy is given by
 $w=-(1/3)\,d\ln(\Omega_m^{-1}-1)/d\ln a$~\cite{Linder:2005in}.
 Using Eq.~(\ref{eq5}) and $E=1/D^\prime$, it becomes
 \bea
 w&=&\frac{2(1+z)EE^\prime-3E^2}{3\left[\,E^2-
 \Omega_{m0}(1+z)^3\,\right]}\label{eq8}\\[1mm]
  &=&-\frac{2(1+z)D^{\prime\prime}+3D^\prime}{3\left[\,D^\prime
  -\Omega_{m0}(1+z)^3 D^{\prime\,3}\,\right]}\,.\label{eq9}
 \eea
 So, $w(z)$ can also be reconstructed from the observational
 $H(z)$ data or SNIa via Gaussian processes.


 \begin{center}
 \begin{figure}[tb]
 \centering
 \vspace{-6mm}  
 \includegraphics[width=0.85\textwidth]{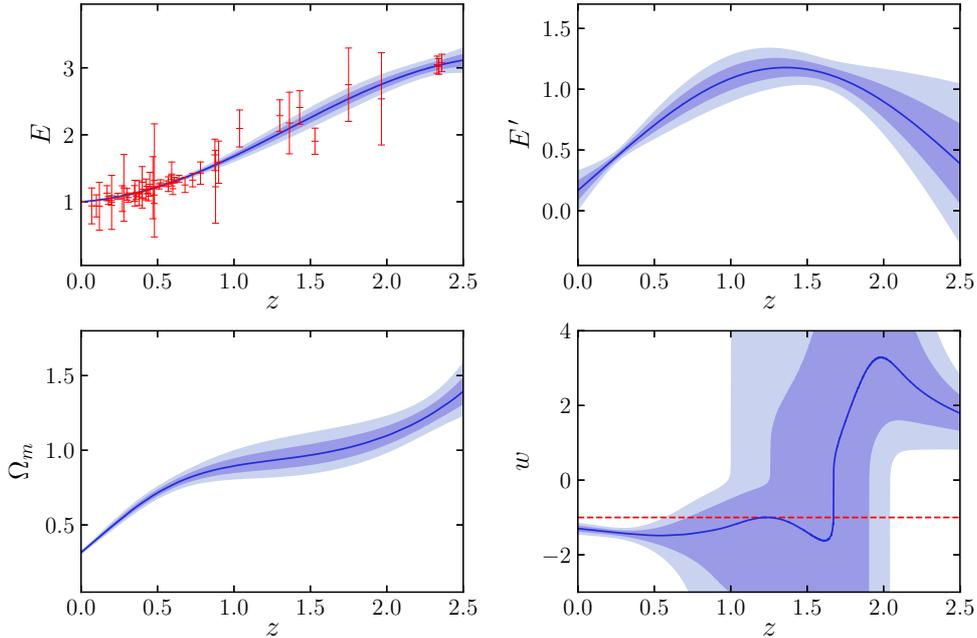}
 \caption{\label{fig4} The same as in Fig.~\ref{fig3}, except
 for the squared exponential covariance function,
 and $H_0=73.52\pm 1.62\,$km/s/Mpc. See the text for details.}
 \end{figure}
 \end{center}



 \begin{center}
 \begin{figure}[p]
 \centering
 \includegraphics[width=0.85\textwidth]{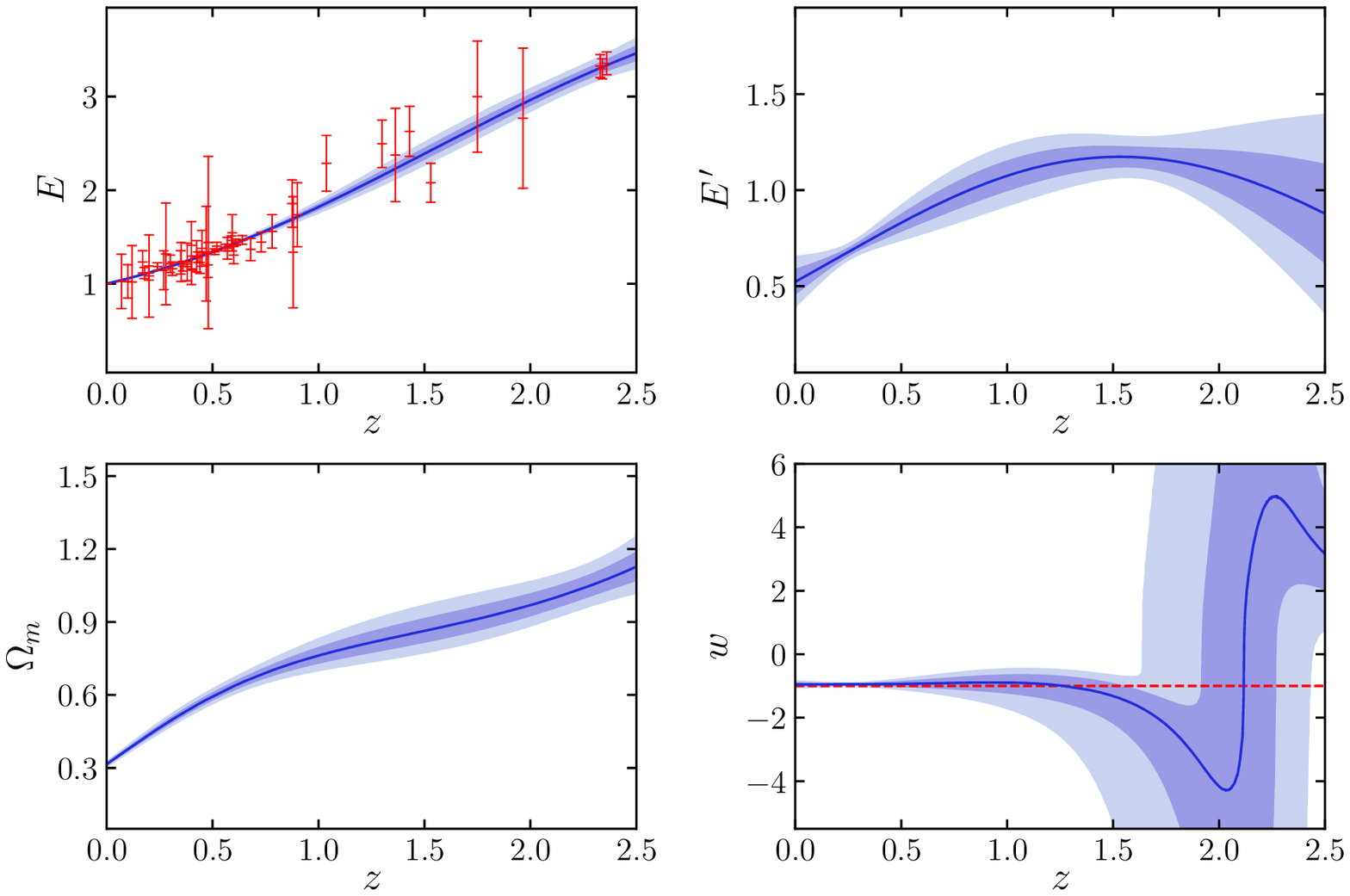}
 \caption{\label{fig5} The same as in Fig.~\ref{fig3}, except
 for the Mat\'ern ($\nu=9/2$) covariance function,
 and $H_0=67.36\pm 0.54\,$km/s/Mpc. See the text for details.}
 \vspace{12mm}  
 \includegraphics[width=0.85\textwidth]{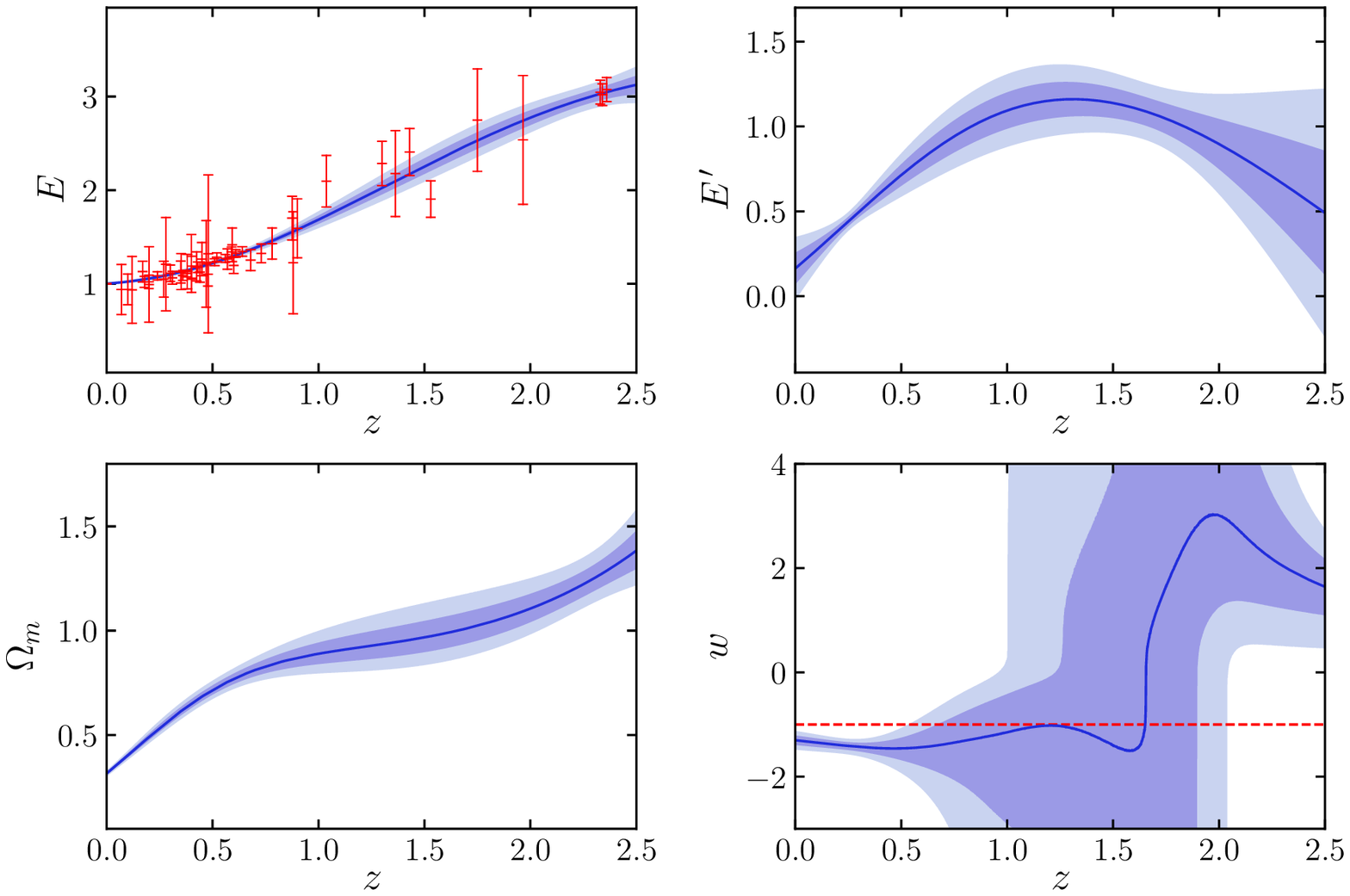}
 \caption{\label{fig6} The same as in Fig.~\ref{fig3}, except
 for the Mat\'ern ($\nu=9/2$) covariance function,
 and $H_0=73.52\pm 1.62\,$km/s/Mpc. See the text for details.}
 \end{figure}
 \end{center}


\vspace{-18mm}  


\section{Results}\label{sec3}

In the following, we use the latest observational data as of
 July 2018. We first reconstruct $f(z)$ via Gaussian processes.
 In~\cite{Kazantzidis:2018rnb}, a sample consisting of 63
 observational $f\sigma_{8,\,obs}$ data published to date are given,
 which is the largest $f\sigma_8$ compilation in the literature
 by now. We can reconstruct $f\sigma_8(z)$ from the observational data,
 and then $\delta^\prime/\delta_0$, $\delta/\delta_0$ and
 $f$ from Eqs.~(\ref{eq3}), (\ref{eq4}) and (\ref{eq2}), respectively.
 Note that in Eqs.~(\ref{eq3}) and (\ref{eq4}), we adopt
 $\sigma_{8,\,0}=0.8111\pm 0.0060$ from the newest Planck
 2018 results~\cite{Aghanim:2018eyx,Akrami:2018vks}. The
 reconstructed $f\sigma_8$, $\delta/\delta_0$,
 $\delta^\prime/\delta_0$ and $f$ as functions of redshift $z$
 are given in Figs.~\ref{fig1} and \ref{fig2}. Clearly, the
 choices of covariance function only make fairly
 small difference.

Then, we reconstruct $\Omega_m(z)$ while $E(z)=H(z)/H_0$ in it
 can be reconstructed by using the expansion history. As
 mentioned above, two ways are viable. At first,
 we consider the 51 observational $H(z)$ data
 compiled in~\cite{Magana:2017nfs}, which is the largest sample
 by now to our best knowledge. Here, the Hubble constant $H_0$
 is required to convert the observational $H(z)$ data into the
 observational $E(z)$ data. As is well known, two observational
 $H_0$ values from the observations at high and low redshifts
 are in significant tension. The newest $H_0=67.36\pm 0.54\,$km/s/Mpc
 from the Planck 2018 results~\cite{Aghanim:2018eyx,Akrami:2018vks}
 is much smaller than the newest $H_0=73.52\pm 1.62\,$km/s/Mpc
 from the SH0ES 2018 results~\cite{Riess:2018byc}, beyond
 $3.6\sigma$ C.L. Since the debate is not settled by now, we
 choose to use them equally. The uncertainties in
 the observational $H(z)$ data and $H_0$ are propagated to the
 $E_{obs}$ data analytically~\cite{Alam:2004ip,Wei:2006va},
 through $\sigma_E^2=\sigma_H^2/H_0^2+(H^2/H_0^4)
 \,\sigma_{H_0}^2$~\cite{Seikel:2012cs}. On the other hand,
 $E(z=0)=1$ exactly by definition. We can reconstruct $E(z)$
 and $E^\prime(z)$ from the observational $E_{obs}$ data via
 Gaussian processes, and then $\Omega_m$ and $w$ from Eqs.~(\ref{eq5})
 and (\ref{eq8}), in which we adopt
 $\Omega_{m0}=0.3153\pm 0.0073$ from the newest Planck 2018
 results~\cite{Aghanim:2018eyx,Akrami:2018vks}. The reconstructed $E$,
 $E^\prime$, $\Omega_m$ and $w$ as functions of redshift $z$
 are given in Figs.~\ref{fig3}$\,\sim\,$\ref{fig6}. Clearly,
 the choices of covariance function only make fairly small
 difference, but the choices of $H_0$ make considerable difference.
 In particular, $w=-1$ is fully consistent with the reconstructed $w(z)$
 in the cases of $H_0=67.36\pm 0.54\,$km/s/Mpc, but $w<-1$ is slightly
 favored in the cases of $H_0=73.52\pm 1.62\,$km/s/Mpc.


 \begin{center}
 \begin{figure}[tb]
 \centering
 \vspace{-4.55mm}  
 \includegraphics[width=0.85\textwidth]{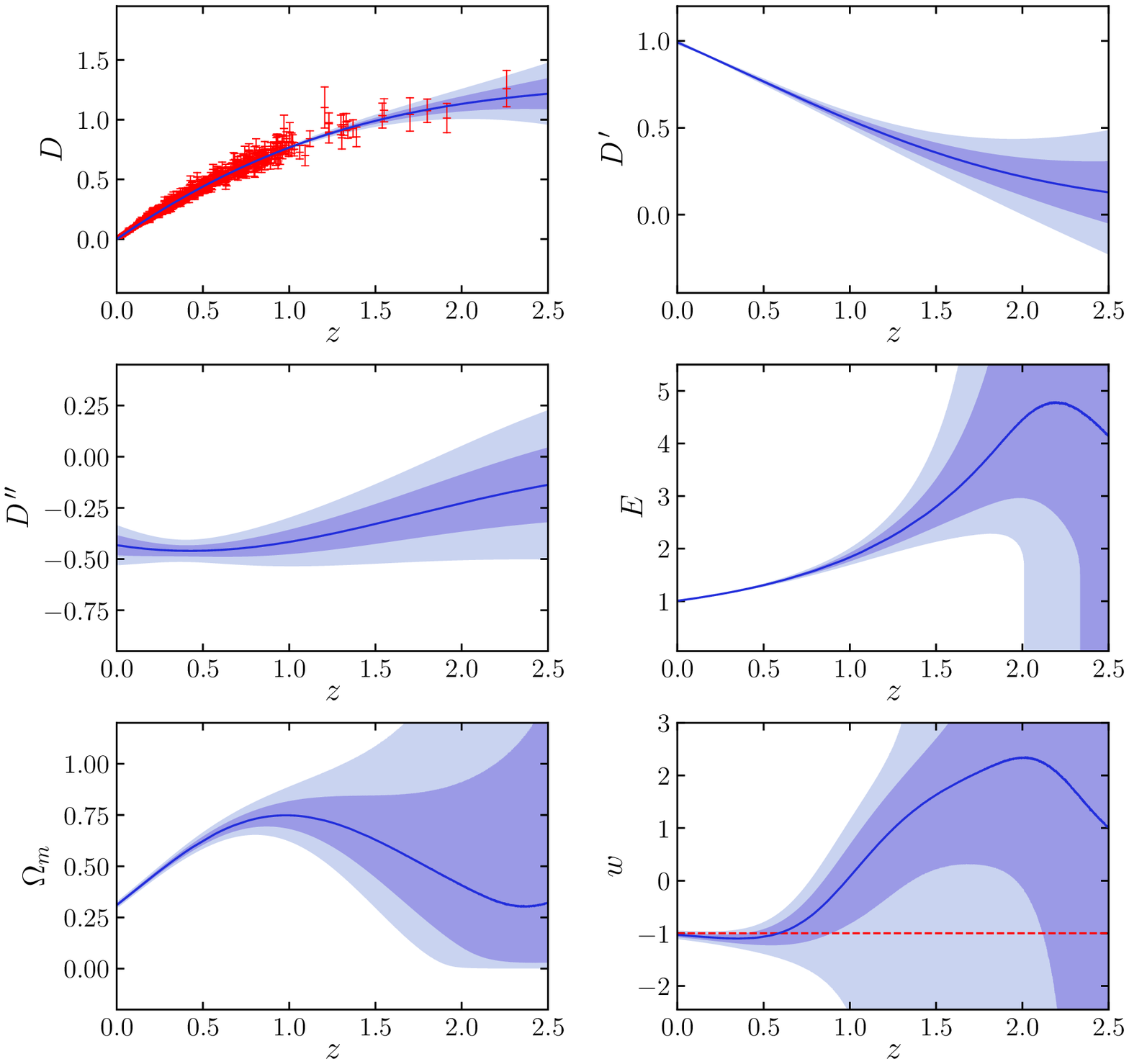}
 \caption{\label{fig7} The reconstructed $D$, $D^\prime$,
 $D^{\prime\prime}$, $E$, $\Omega_m$ and $w$ as functions of
 redshift $z$, by using Gaussian processes with the squared
 exponential covariance function, from the Pantheon SNIa data.
 The mean and $1\sigma$, $2\sigma$ uncertainties are indicated
 by the blue solid lines and the shaded regions, respectively.
 The observational $D_{obs}$ data with error bars are also
 plotted in the top-left panel. $w=-1$ is indicated by a
 red dashed line. See~the~text~for~details.}
 \end{figure}
 \end{center}



 \begin{center}
 \begin{figure}[tb]
 \centering
 \vspace{-6mm}  
 \includegraphics[width=0.85\textwidth]{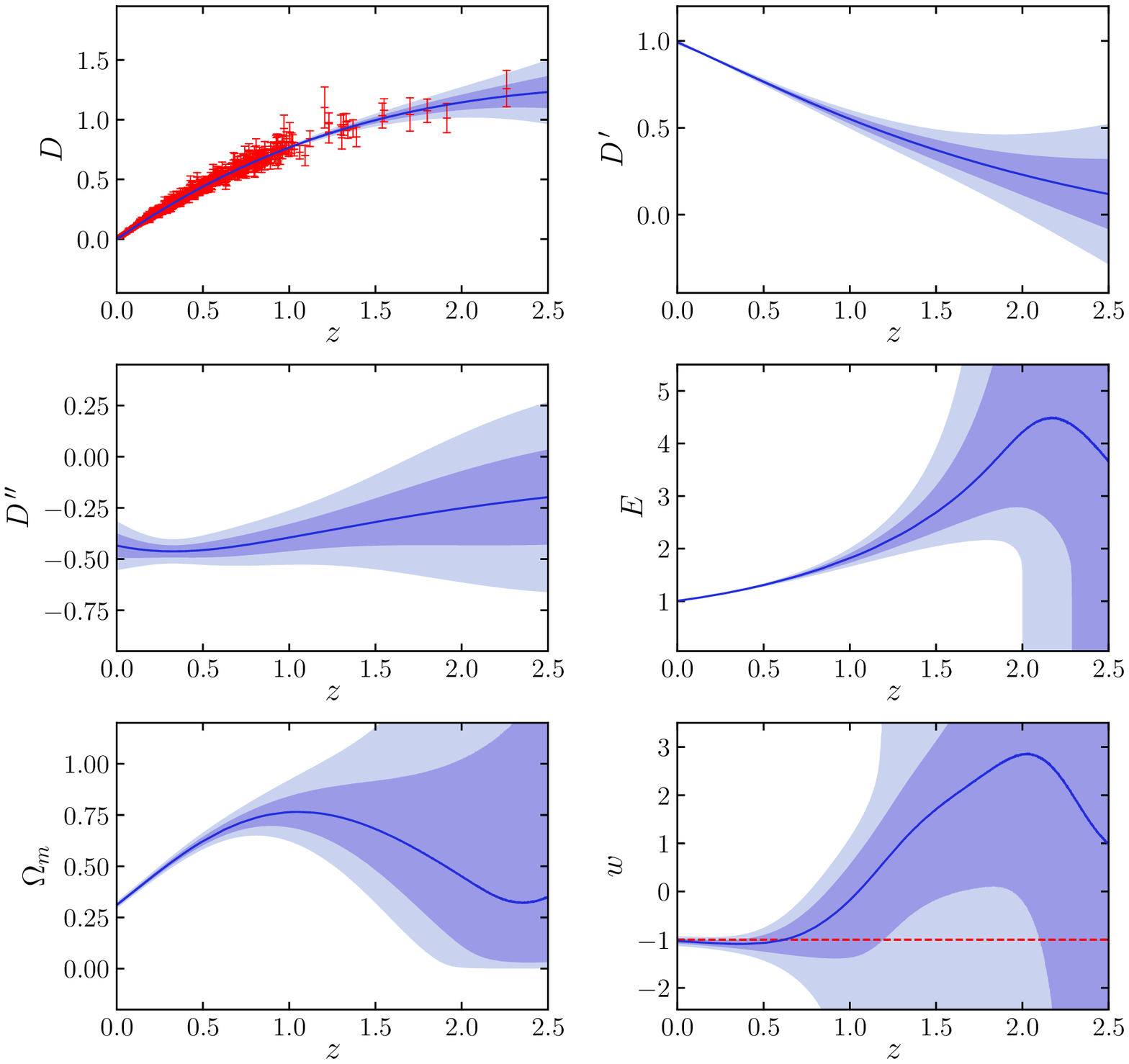}
 \caption{\label{fig8} The same as in Fig.~\ref{fig7}, except
 for the Mat\'ern ($\nu=9/2$) covariance function. See the
 text for details.}
 \end{figure}
 \end{center}


\vspace{-18.5mm} 

As an alternative, we can also reconstruct $E=1/D^\prime$ from
 the luminosity distance of SNIa. Here we use the Pantheon
 sample~\cite{Scolnic:2017caz,Pantheondata,Pantheonplugin}
 consisting of 1048 SNIa, which is the largest
 spectroscopically confirmed SNIa sample by now. The corrected
 bolometric apparent magnitude $m$ is related to $D$ according
 to~\cite{Deng:2018jrp}
 \be{eq10}
 m = 5\log_{10}\left(\left(1+z\right)D\right)+{\cal M}\,,
 \ee
 where $\cal M$ is a nuisance parameter representing some
 combination of the absolute magnitude $M$ and $H_0$. One can
 convert the observational $m$ data given in the
 Pantheon plugin~\cite{Pantheonplugin} into the $D_{obs}$
 data, while their covariance matrices are related by the
 propagation of uncertainty~\cite{poucov},
 $\boldsymbol{C}_D=\boldsymbol{J}\boldsymbol{C}_m \boldsymbol{J}^{\,T}$,
 where $\boldsymbol{J}$ is the Jacobian matrix. Note that we consider
 the full covariance matrix with the systematic uncertainties.
 Here we adopt the best-fit ${\cal M}=23.808891$ for the flat
 $\Lambda$CDM model~\cite{Deng:2018jrp} as a fiducial value. We can
 reconstruct $D(z)$, $D^\prime(z)$ and $D^{\prime\prime}(z)$
 from the observational $D_{obs}$ data via Gaussian processes,
 and then $E=1/D^\prime$, as well as $\Omega_m$ and $w$ from
 Eqs.~(\ref{eq5}) and (\ref{eq9}), in which we adopt again
 $\Omega_{m0}=0.3153\pm 0.0073$ from the newest Planck 2018
 results~\cite{Aghanim:2018eyx,Akrami:2018vks}.
 The reconstructed $D$, $D^\prime$, $D^{\prime\prime}$, $E$,
 $\Omega_m$ and $w$ as functions of redshift $z$ are given in
 Figs.~\ref{fig7} and \ref{fig8}. Clearly, the choices of
 covariance function only make fairly small difference. In
 both cases, $w=-1$ is fully consistent with the reconstructed $w(z)$.

\begin{center}
 \begin{figure}[tb]
 \centering
 \vspace{-3mm}  
 \includegraphics[width=0.85\textwidth]{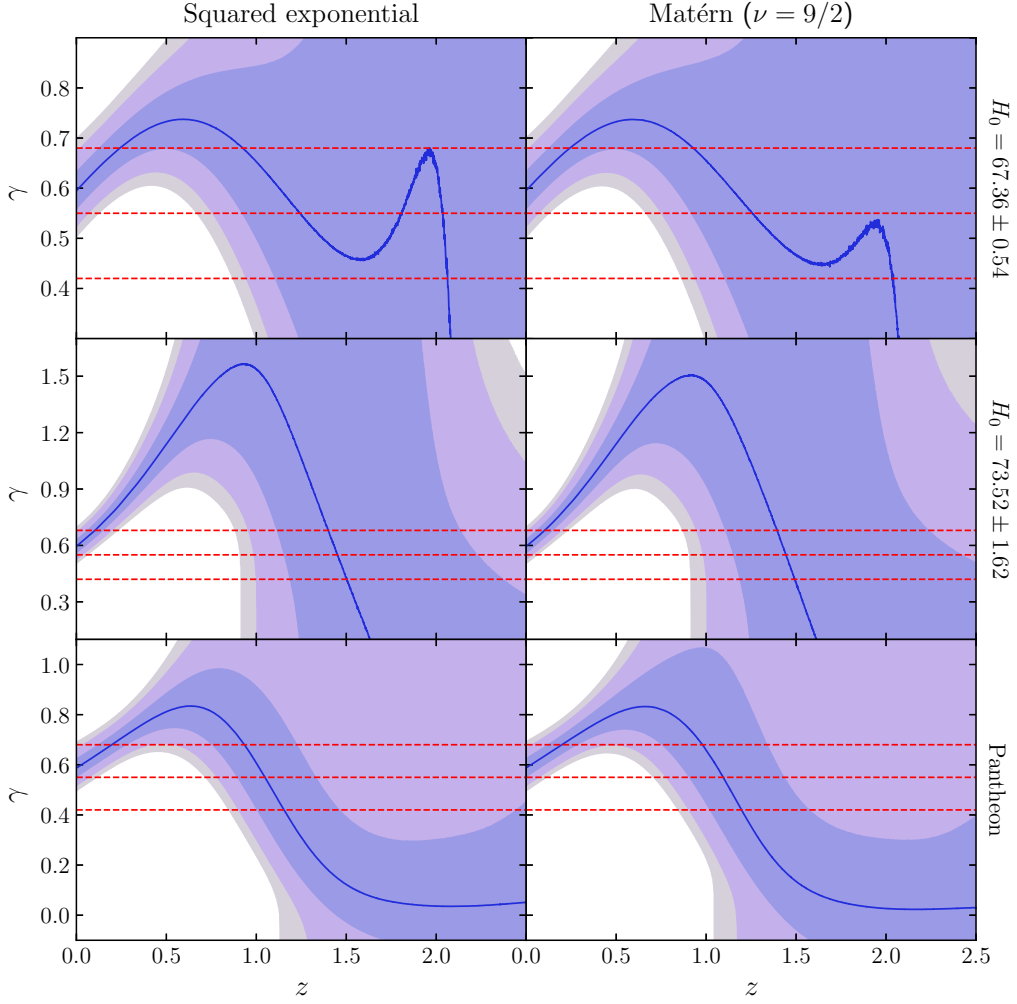}
 \caption{\label{fig9} The reconstructed growth index $\gamma$ as a
 function of redshift $z$, by using the latest observational data via
 Gaussian processes. The mean and $1\sigma$, $2\sigma$,
 $3\sigma$ uncertainties are indicated by the blue solid lines
 and the shaded regions, respectively. The left and right
 panels correspond to the squared exponential and the Mat\'ern
 ($\nu=9/2$) covariance functions, respectively. The top,
 middle, and bottom panels correspond to the observational
 $H(z)$ data with $H_0=67.36\pm 0.54\,$km/s/Mpc,
 $H_0=73.52\pm 1.62\,$km/s/Mpc, and the Pantheon SNIa data,
 respectively. $\gamma=0.42$ ($f(R)$
 theories), $0.55$ (dark energy models in GR, especially
 $\Lambda$CDM), $0.68$ (DGP model) are indicated by the red
 dashed lines. See the text for details.}
 \end{figure}
 \end{center}

Finally, using the above reconstructed $f(z)$ and $\Omega_m(z)$, we
 obtain the growth index $\gamma$ as a function of redshift $z$
 from Eq.~(\ref{eq1}). We present the reconstructed $\gamma(z)$
 in Fig.~\ref{fig9}, for various observational data and Gaussian
 processes with different covariance functions. It is easy to
 see that the choices of covariance function only make fairly
 small difference. Notably, $\gamma\simeq 0.42$ ($f(R)$
 theories)~\cite{Gannouji:2008wt,Tsujikawa:2009ku,Shafieloo:2012ms,
 Tsujikawa:2010zza} is inconsistent with the reconstructed
 $\gamma(z)$ in the redshift range $0\leq z\;\lsim\; 0.8$ far beyond
 $3\sigma$ C.L. in all cases (see Fig.~\ref{fig9}). Also,
 $\gamma_0\simeq 0.42$ at $z=0$ is strongly disfavored at very
 high C.L. On the other hand, also in all cases, although
 $\gamma_0\simeq 0.55$ at $z=0$ and $\gamma\simeq 0.55$ (dark energy
 models within GR)~\cite{Linder:2005in} at low redshift
 $z\;\lsim\; 0.1$ are consistent with the results, from
 Fig.~\ref{fig9} one can see that $\gamma\simeq 0.55$ is still
 inconsistent with the reconstructed $\gamma(z)$ in the
 moderate redshift range $0.1\;\lsim\; z\;\lsim\; 0.7$ far
 beyond $3\sigma$ C.L., due to the arched structure in the
 reconstructed $\gamma(z)$. In particular, for $\Lambda$CDM
 model $\gamma=6/11\simeq 0.545$~\cite{Linder:2005in,Linder:2007hg},
 which is approximately independent of redshift. Therefore, it
 is also disfavored far beyond $3\sigma$ C.L. At last, the
 status is fairly subtle for $\gamma\simeq 0.68$
 (DGP model)~\cite{Linder:2007hg,Wei:2008ig}. From Fig.~\ref{fig9},
 we find that in the cases of the observational $H(z)$ data
 with $H_0=67.36\pm 0.54\,$km/s/Mpc and the Pantheon SNIa data,
 $\gamma\simeq 0.68$ is fully consistent with the reconstructed
 $\gamma(z)$. However, in the case of the observational $H(z)$ data
 with $H_0=73.52\pm 1.62\,$km/s/Mpc, $\gamma\simeq 0.68$ is
 inconsistent with the results in the moderate redshift range
 (see the middle panels of Fig.~\ref{fig9}), due to the arched
 structure in the reconstructed $\gamma(z)$. Since the tension
 between the above two $H_0$ is beyond $3.6\sigma$, and the
 debate in the community is not settled by now, we can say
 nothing certainly about $\gamma\simeq 0.68$ (DGP model)
 so far.

Nevertheless, our results strongly suggest that the growth
 index $\gamma(z)$ is varying. The arched structure in the
 reconstructed $\gamma(z)$ shown in Fig.~\ref{fig9} plays
 an important role. The derivative $\gamma^\prime >0$ at low
 redshift, and $\gamma^\prime <0$ at higher redshift. The
 models with $\gamma\;\lsim\; 0.55$ in the moderate redshift
 range $0.1\;\lsim\; z\;\lsim\; 0.7$ are disfavored far
 beyond $3\sigma$ C.L.


\vspace{-15mm}  


\section{Conclusion and discussion}\label{sec4}

The accelerated cosmic expansion could be due to dark energy
 within GR, or modified gravity. It is of interest to differentiate
 between them, by using both the expansion history and the
 growth history. In the literature, it was proposed that the
 growth index $\gamma$ is useful to distinguish these two scenarios.
 In this work, we consider the non-parametric reconstruction of
 the growth index $\gamma$ as a function of redshift~$z$ from
 the latest observational data as of July 2018 via Gaussian
 Processes. Interestingly, we find that $f(R)$ theories and
 dark energy models within GR (especially $\Lambda$CDM) are
 inconsistent with the results in the moderate redshift range
 far beyond $3\sigma$ C.L., due to the arched structure in the
 reconstructed $\gamma(z)$. A modified gravity scenario
 different from $f(R)$ theories is favored.

Obviously, this result is unusual, and new physics is required.
 However, it does not mean that dark energy models within GR
 (especially $\Lambda$CDM) and $f(R)$ theories certainly end.
 First, one can doubt the observational data used in this work.
 For instance, the 63 observational $f\sigma_{8,\,obs}$ data
 compiled in~\cite{Kazantzidis:2018rnb} might be correlated,
 and contain duplicated points from the same surveys, while the
 corrections from the choices of the fiducial cosmology should
 be taken into account. So, this $f\sigma_{8,\,obs}$ sample
 might require a re-analysis, as preliminarily considered
 in~\cite{Nesseris:2017vor,Kazantzidis:2018rnb}. Second, the
 reliability of Gaussian Processes at high redshift might be
 questionable. Therefore, we should test the growth index
 $\gamma$ by using another independent method, and cross-check
 the corresponding results with the ones from Gaussian
 Processes. In fact, our relevant work will appear in a
 separate paper~\cite{Yin:2019rgm}. Finally, in the present
 work, cold dark matter is implicitly assumed, as in Eq.~(\ref{eq5}).
 If there is a non-zero interaction between dark energy
 and dark matter, Eq.~(\ref{eq5}) should be changed to
 \be{eq11}
 \Omega_m(z)\equiv\frac{8\pi G\rho_m}{3H^2}=
 \frac{\Omega_{m0}(1+z)^{3+\xi}}{E^2(z)}\,,
 \ee
 where $\xi$ characterizes the deviation from uncoupled cold
 dark matter (note that $\xi$ can be time-dependent in
 general). Another non-trivial possibility assumes that dark
 matter is not cold. In fact, for warm dark matter, its
 equation-of-state parameter $w_m\not=0$, and
 hence $\Omega_m(z)$ takes a form similar to Eq.~(\ref{eq11}).
 In both non-trivial cases, the conclusions should be changed,
 and dark energy models within GR (especially $\Lambda$CDM) and
 $f(R)$ theories might still survive (see~\cite{Wei:2008vw,
 Wei:2013rea} for deeper discussions). The new physics in these
 non-trivial cases lies in the non-zero interaction between
 dark energy and dark matter, or the induction of warm dark
 matter. They deserve further investigations.

After all, we would like to mention several technical details.
 One might note that in Figs.~\ref{fig3}$\,\sim\,$\ref{fig6}
 the reconstructed $\Omega_m$ becomes larger than $1$ at high
 redshift $z\,\gsim\, 2$, but this is not unphysical in fact.
 Yes, in GR, the Friedmann equation
 $H^2=8\pi G(\rho_m+\rho_X)/3$ is unchanged, and hence
 $0\leq\Omega_m\leq 1$ by definition. However, this is right
 only for dark energy models in GR. On the contrary, in
 modified gravity, the Friedmann equation should be modified,
 and the modification to GR can be equivalent to an effective
 ``energy component''. So, the effective $\rho_X$ can be
 negative at high redshift, and hence $\Omega_m>1$ is possible.
 Since the effective ``energy component'' is not real matter
 (it is actually the modification to GR, namely a geometric
 effect indeed), this is allowed by physics. In fact, noting
 that $0\leq\Omega_m\leq 1$ must be held for dark energy
 models in GR, our reconstructed $\Omega_m>1$ at high redshift
 $z\,\gsim\, 2$ in Figs.~\ref{fig3}$\,\sim\,$\ref{fig6} can be
 regarded as an extra evidence supporting modified gravity
 against dark energy models in GR.

It is known that in modified gravity, e.g. $f(R)$ theories, the
 growth rate $f=f(z,k)$ is also spatially scale-dependent in general
 (see e.g.~\cite{Gannouji:2008wt,Tsujikawa:2009ku,Jennings:2012pt}),
 where the comoving wavenumber $k$ denotes the
 scale~\cite{Tsujikawa:2007gd}. In principle, the
 scale-dependence should be taken into account (we thank the referee
 for pointing out this issue). However, let us have a closer
 look. In~\cite{Gannouji:2008wt}, they found that $f$ and hence
 $\gamma$ is scale-independent at redshift $z\,\lsim\, 0.5$
 (see their Fig.~2 and the text below Eq.~(4.17) or Eq.~(58) in
 the arXiv version). At $z=0$, they found $\gamma_0\simeq 0.41$
 independent of the scales $k$. At higher redshift, they have a
 small difference $\Delta\gamma\,\lsim\, 0.04$ between various
 scales. $\gamma$ becomes smaller as redshift $z$ increases, so
 that $\gamma\,\lsim\, 0.41$ at higher redshift.
 In~\cite{Tsujikawa:2009ku}, the results are quite similar.
 They found that the dispersion of $\gamma$ with respect to
 the scale $k$ is very small (see their Fig.~2 and the text in
 Sec.~IV.B), namely $\gamma$ is nearly scale-independent at
 redshift $z\,\lsim\, 1$ (especially the dispersion of $\gamma$
 is nearly absent for scales $k\geq 0.033h\,{\rm Mpc}^{-1}$).
 Again, $\gamma$ becomes smaller as redshift $z$ increases. On
 the other hand, from their Figs.~1, 3, 4, 6, 7, one can see
 that $\gamma_0=\gamma(z=0)$ is smaller than $\sim 0.557$ for
 various model parameters of viable $f(R)$ theories, and this
 is independent of the scales $k$. Keeping the above results
 of~\cite{Gannouji:2008wt,Tsujikawa:2009ku} in mind, let us
 turn back to our Fig.~\ref{fig9}. First, in all cases
 of Fig.~\ref{fig9}, $\gamma_0\,\lsim\, 0.5$ at $z=0$ is
 clearly inconsistent with our reconstructed $\gamma(z)$ beyond
 $3\sigma$ C.L. As mentioned above, it is found
 in~\cite{Gannouji:2008wt,Tsujikawa:2009ku} that
 $\gamma_0=\gamma(z=0)$ is smaller than $\sim 0.557$ for
 various model parameters of viable $f(R)$ theories, and this
 is independent of the scales $k$. So, our results of $\gamma_0$ is
 a bad news to most of these viable $f(R)$ theories, although it is
 not so decisive. Second, in all cases of Fig.~\ref{fig9},
 $\gamma<0.56$ in the moderate redshift range
 $0.1\,\lsim\, z\,\lsim\, 0.7$ is also clearly inconsistent with our
 reconstructed $\gamma(z)$ far beyond $3\sigma$ C.L. Actually,
 in all cases of Fig.~\ref{fig9}, even $\gamma\,\lsim\, 0.6$ is
 still inconsistent with our reconstructed $\gamma(z)$ beyond
 $3\sigma$ C.L. in a relatively narrower moderate redshift range. As
 mentioned above, it is found
 in~\cite{Gannouji:2008wt,Tsujikawa:2009ku} that $\gamma$
 becomes smaller as redshift $z$ increases. Together with the
 fact that $\gamma_0=\gamma(z=0)$ is smaller than $\sim 0.557$
 for various model parameters of viable $f(R)$ theories, it is
 easy to see that $\gamma\,\lsim\, 0.557<0.56$ in the moderate
 redshift range $0.1\,\lsim\, z\,\lsim\, 0.7$. This is also
 independent of the scales $k$. Note that we can further relax
 $0.56$ to the larger $0.6$ as mentioned above. Therefore, we
 can still say that most of viable $f(R)$ theories with various
 model parameters are inconsistent with our reconstructed
 $\gamma(z)$ in the moderate redshift range beyond $3\sigma$
 C.L., and this conclusion is nearly scale-independent
 actually. Finally, even in the worst case that our results are
 not applicable to $f(R)$ theories due to the scale-dependence,
 our other conclusion that dark energy models in GR (especially
 $\Lambda$CDM) are inconsistent with our reconstructed
 $\gamma(z)$ in the moderate redshift range far beyond
 $3\sigma$ C.L. is still valid. New physics is still required.


\section*{ACKNOWLEDGEMENTS}

We thank the anonymous referee for useful comments and
 suggestions, which helped us to improve this work. We are
 grateful to Hua-Kai~Deng, Xiao-Bo~Zou, Da-Chun~Qiang,
 Zhong-Xi~Yu, and Shou-Long~Li for kind help and
 discussions. This work was supported in part by
 NSFC under Grants No.~11575022 and No.~11175016.

\renewcommand{\baselinestretch}{1.0}


\end{document}